\documentclass[journal=jpccck,manuscript=article,layout=twocolumn]{achemso}

\usepackage[utf8]{inputenc}
\usepackage[T1]{fontenc} 
\usepackage{amsmath}
\usepackage{amssymb}
\usepackage{lineno}
\usepackage{color}

\author{Klaus H. Eckstein}
\affiliation[uniwue]{Institute of Physical and Theoretical Chemistry, Julius-Maximilian University W\"urzburg, Germany}
\phone{+49 931 3188969}
\email{klaus.eckstein@uni-wuerzburg.de}
\author{Florian Hirsch}
\affiliation[uniwue]{Institute of Physical and Theoretical Chemistry, Julius-Maximilian University W\"urzburg, Germany}
\author{Richard Martel}
\affiliation[umontreal]{Department of Chemistry, University of Montreal, Canada}
\author{Tobias Hertel}
\affiliation[uniwue]{Institute of Physical and Theoretical Chemistry, Julius-Maximilian University W\"urzburg, Germany}
\alsoaffiliation{R\"ontgen Research Center for Complex Material Systems, Julius-Maximilian University W\"urzburg, Germany}

\title {Infrared Study of Charge Carrier Confinement in Doped (6,5) Carbon Nanotubes}


\begin{document}

\begin{abstract}
Electronic degrees of freedom and their coupling to lattice vibrations in semiconductors can be strongly modified by doping. Accordingly, the addition of surplus charge carriers to chirality-mixed carbon nanotube samples has previously been found to give rise to a Drude-type plasmon feature as well as Fano-type antiresonances in the far- to mid-infrared spectral range (FIR/MIR). Here we investigate the FIR/MIR response of redox-chemically doped semiconducting (6,5) carbon nanotubes (s-SWNTs). We find that, contrary to expectations, the Drude-type plasmon shifts to {\it lower} wavenumbers with increasing doping level. By means of Monte-Carlo simulations of the optical response, we attribute this behavior to the confinement of excess charge carriers at low doping levels and their progressive delocalization when approaching degenerate doping. The coupling of vibrational modes to intraband excitations in the doped s-SWNTs can be probed via a double resonance process similar to that responsible for the Raman D-band. The resulting Fano antiresonances shed new light onto the character and coupling of electronic and vibrational degrees of freedom in these one-dimensional semiconductors.
\end{abstract}


\section{Introduction}

Plasmon and Fano resonances in carbon nanotubes have been studied extensively to better understand how these spectroscopic signatures may be used for accessing microscopic aspects of the electronic degrees of freedom in these systems as well as for probing how electronic and vibrational degrees of freedom couple with one another \cite{Bermudez2006, Nakanishi2009, Lapointe2012, Zhang2013, Morimoto2014, Shi2015, Lapointe2017, Chiu2017, Zhukova2017}.  Plasmonic properties of carbon nanotubes and other nano-materials are also explored for their practical use, for example in surface enhanced infrared absorption \cite{Yang2018} or infrared detectors \cite{MahjouriSamani2013}. 

Doping plays a decisive role in securing and tailoring the functionality of any such semiconductor-based spectroscopic or opto-electronic applications. However, the understanding of how concentration and the distribution of excess charge carriers in semiconducting single-wall carbon nanotubes (s-SWNTs) can be controlled or monitored is still evolving \cite{Zorn2020, Ferguson2018, Eckstein2017, Eckstein2019}. Far- and mid-infrared spectroscopy (FIR/MIR) here represents a versatile tool for probing the electronic and vibrational degrees of freedom in such nanoscale structures.

FIR to MIR spectra of mixed-chirality nanotube samples are dominated by a broad absorption band, which is commonly attributed to a plasmon resonance \cite{Zhang2013, Morimoto2014, Chiu2017, Zhukova2017}. The frequency of this resonance $\nu_p$ is predicted to depend on the nanotube length $l$ as $\nu_p\propto v_F/l$, where $v_F$ is the Fermi velocity \cite{Nakanishi2009, Morimoto2014}. The dependence on nanotube length appears to have been confirmed \cite{Zhang2013, Morimoto2014, Chiu2017} but the effect of dopant concentration appears to have eluded closer experimental inspection, possibly because reliable protocols for attaining and quantifying doping levels have become available only recently \cite{Eckstein2019}. 

Fano antiresonances in the MIR spectrum of mixed-chirality carbon nanotubes and monolayer graphene have also shed some light on the coupling of low-energy intraband electronic excitations to phonons, \cite{Lapointe2012, Lapointe2017, Chiu2017} similar to the Breit-Wigner-Fano lineshapes observed in resonance Raman spectra of metallic nanotubes \cite{Brown2001, Wu2007, Hasdeo2013}. Specifically, the intensity of MIR Fano resonances was found to increase with both, doping level and defect density. The proposed mechanism is based on a combination of inelastic phonon and elastic defect scattering \cite{Lapointe2012, Lapointe2017}.

What remains somewhat unclear is how doping level and sample character may affect optical spectra. The majority of prior FIR/MIR studies was performed on samples consisting of mixtures of metallic and semiconducting SWNTs or made of larger diameter semiconducting SWNTs with different chiralities \cite{Zhang2013, Morimoto2014}. Moreover, previous interpretations of MIR plasmon features in doped SWNTs were often based on the implicit assumption that the excess charge distribution is homogeneous, which may not always be the case \cite{Lapointe2012, Eckstein2017, Eckstein2019}. In addition, information on doping for plasmon related optical studies has been mostly qualitative rather than quantitative \cite{Eckstein2019}.

The work presented here thus discusses broadband transmission spectra from highly chirality enriched (6,5)-SWNTs from the UV to the far infrared over a broad range of quantifiable carrier concentrations. Upon doping, we observe pronounced simultaneous changes of the inter- and intraband transitions. Spectral shifts in the mid- and far-infrared are analyzed with regards to the degree of carrier confinement using Monte-Carlo simulations of the intraband optical conductivity. In addition, we analyze the dependence of MIR absorption Fano antiresonances in the same spectral range as the well known Raman G- and D-bands. These bands are particularly intriguing as their appearance, strength and character is found to be intimately tied to the Drude-type plasmon, the phonon spectrum, as well as a double resonance facilitated by defect scattering similar to that of the graphene and nanotube Raman D-band.   

\section{Experimental Section}

\textbf{Sample preparation}

Single-wall carbon nanotube suspensions of the (6,5) species in toluene solvent (analytical reagent grade, Fischer Chemicals) were fabricated by shear force mixing 0.38 mg/mL CoMoCAT raw material (SWeNT SG 65i, Southwest Nano Technologies Inc.) with 0.76 mg/mL PFO-BPy (American Dye Source) for 12 h, following the process described by Graf et al.~\cite{Graf2016}. Subsequent centrifugation and extraction of the supernatant yields dispersions with high purity of semiconducting SWNTs (> 99.98\%~\cite{Brady2016}) that are also highly enriched with the (6,5)-chirality (> 95\% based on absorption spectroscopy). 

Unintentional doping of as-prepared s-SWNT samples appears to be negligible as evidenced by the absence of a clear trion signal or an IR, Drude-like absorption band (see Figure \ref{fig1}a). The optical properties of the non-doped sample are thus characteristic of the intrinsic semiconductor material.

SWNT thin-films were prepared by vacuum filtration of the SWNT dispersion through cellulose acetate filter membranes (MF-Millipore VCWP, Merck Millipore) with 100\,nm pore size. The filter membrane was dissolved in an acetone bath and freely floating SWNT films were deposited on 500\,$\mu$m thick ultra-high molecular weight polyethylene substrates (Goodfellow).

SWNT films were doped by immersion into solutions (volume ratio toluene:acetonitrile 5:1) of gold(III) chloride (AuCl$_3$, >\,99.99\% purity, Sigma-Aldrich) or triethyloxonium hexachloroantimonate (OA, Sigma-Aldrich) for 10 minutes and subsequent drying of the films. The doping level was controlled via the oxidant concentration. Both of these strong oxidants resulted in quantitatively similar nanotube spectra.\newline

\textbf{Broadband transmission spectroscopy}

Spectra from the far infrared to the UV were recorded using three different spectrometers. UV-VIS-NIR absorption measurements were carried out with a Cary-5000 (Agilent) absorption spectrophotometer. Infrared spectra were recorded with a resolution of $16\rm\, cm^{-1}$ using a Jasco FT/IR-4100 (MIR region, low-wavenumber cutoff at $400\,\rm cm^{-1}$) and with a Bruker IFS120HR spectrometer (FIR/MIR region) when the low-wavenumber cutoff was to be extended to $50\,\rm cm^{-1}$. All spectra were background corrected using a pristine polyethylene substrate as reference.

\section{Results and Discussion}

\subsection{Plasmon resonances}
\begin{figure}[htbp]
	\centering
		\includegraphics[width=8.5 cm]{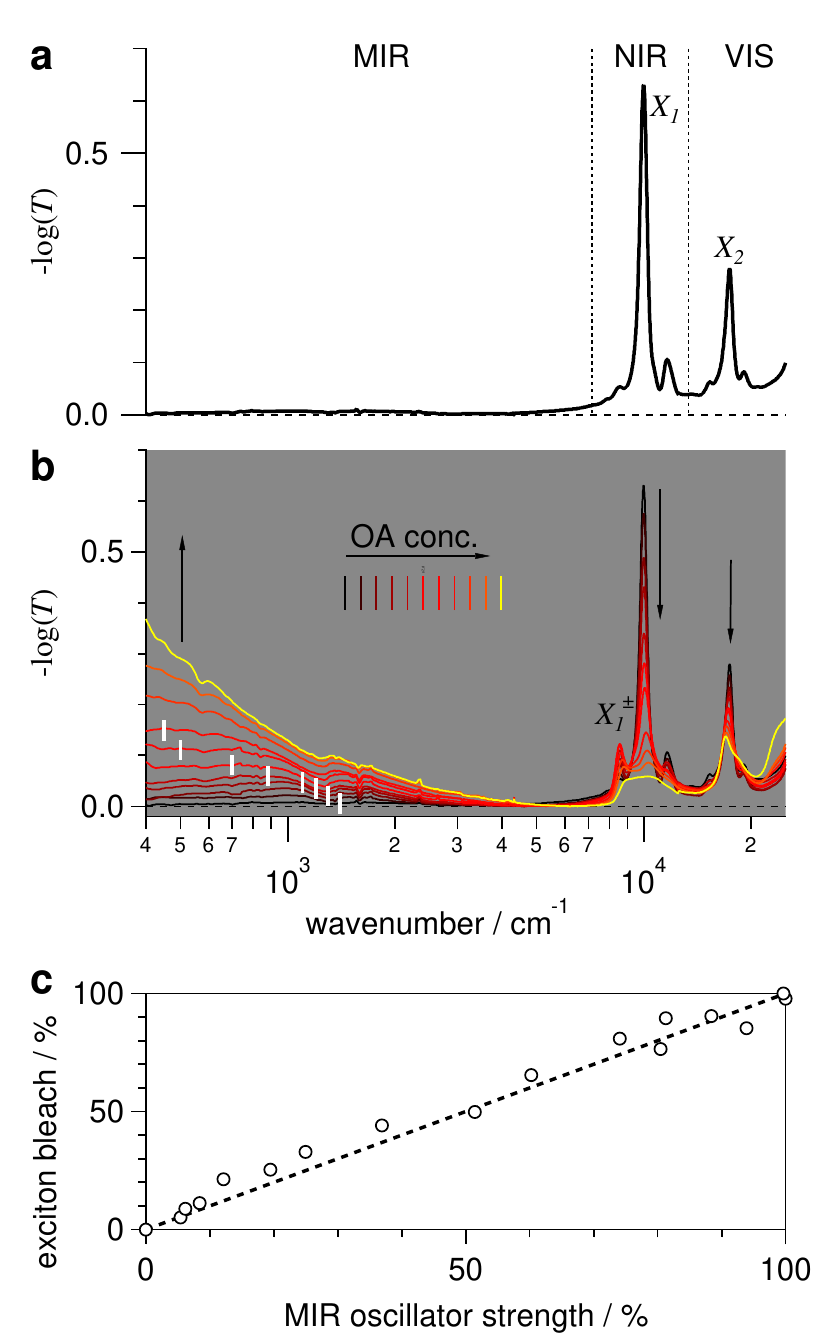}
		\caption{{\bf Effect of doping on the optical spectra of (6,5) s-SWNTs.} {\bf a)} Spectrum of an intrinsic nanotube film. {\bf b)} The same film after treatment with triethyloxonium hexachloroantimonate (OA) hole dopant. The approximate maximum of the broad MIR absorption feature is indicated for the low to moderately doped samples by white vertical bars. {\bf c)} Comparison of the $X_1$ exciton bleach with the increase of MIR oscillator strength.}
		\label{fig1}
\end{figure}

The absorption spectrum of an as-prepared (6,5)-SWNT film is shown in Figure \ref{fig1}a. Its NIR and VIS spectral regions are dominated by the first and second subband exciton absorption, which are here designated as $X_1$- and $X_2$-bands at $10{,}000\,\rm cm^{-1}$ and $17{,}400\,\rm cm^{-1}$, respectively. 

Doping of semiconductors with direct band gap is typically evidenced by the emergence of an absorption band associated with a charged exciton $X_1^\pm$ (trion), which is stabilized with respect to the transition energy of the neutral $X_1$ exciton. Additional signatures of increasing doping levels are a bleach of excitonic transitions as well as the increase of absorbance in the mid- to far-infrared spectral range, all of which have been observed in doped s-SWNTs \cite{Santos2011, Matsunaga2011, Hartleb2015, Avery2016, Zheng2004, Lapointe2012, Zhang2013}.

The spectra of triethyloxonium hexachloroantimonate (OA) {\it p}-doped (6,5) s-SWNTs in Figure \ref{fig1}b exhibit all of the aforementioned characteristics. The $X^+$ band is here found to be centered at $8{,}580\,\rm cm^{-1}$ and Figure \ref{fig1}c reveals that the bleach of the first subband exciton transition scales linearly with the increase of the oscillator strength in the experimentally accessible MIR range from 400 to $4{,}840\,\rm cm^{-1}$. This suggests that spectral changes in the MIR and in the NIR may be used equally well as quantitative predictors of nanotube doping. Details on how the decrease of the exciton oscillator strength can be used to quantify doping levels in s-SWNTs were recently discussed by Eckstein {\it et al.} \cite{Eckstein2019}.

The MIR absorption signal of weakly doped samples in Figure \ref{fig1}b is dominated by a broad band with its peak slightly above $1{,}000$ wavenumbers. Similar absorption features were also found in a variety of mixed metallic and doped semiconducting SWNT samples, see for example references \citep{Zhang2013, Morimoto2014, Chiu2017}. For metallic SWNTs the frequency of the 1D free-carrier plasmon was predicted to be inversely proportional to the tube length \cite{Nakanishi2009}. Similarly, the frequency of free carrier plasmons in doped semiconductors is expected to also increase with doping level. The observation of such MIR resonances has thus frequently been attributed to 1D plasmon excitations in these systems \cite{Zhang2013, Morimoto2014, Chiu2017}. In the doped nanotube samples studied here, however, this feature is seen to shift to {\it lower} frequencies with increasing nanotube doping level, as indicated by the vertical white bars in Figure \ref{fig1}b. This is intriguing and will be analyzed in further detail below. 

\begin{figure}[htbp]
	\centering
		\includegraphics[width=8.5 cm]{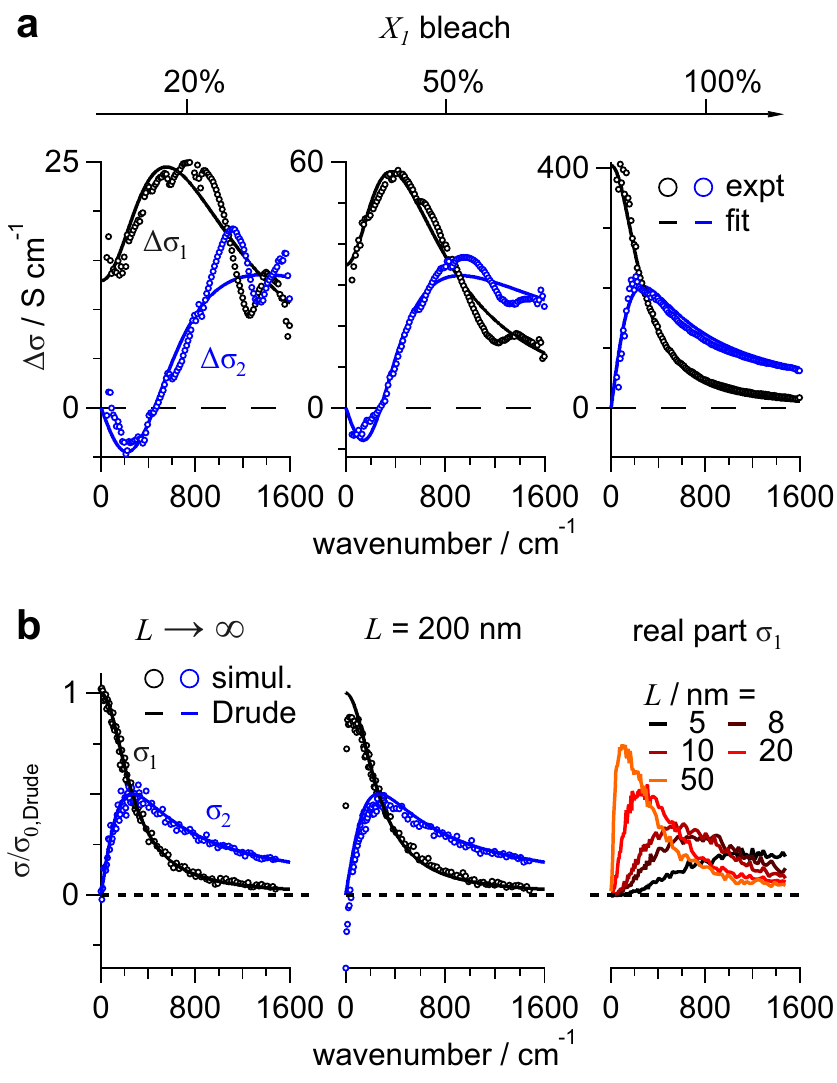}
		\caption{{\bf a)} Drude-Smith fit to real and imaginary parts of the optical conductivities for low and moderate doping levels (left panels) and a Drude fit for a degenerately doped system (right panel). {\bf b)} Monte-Carlo simulations of the optical conductivity for very large charge confinement lengths, yields the Drude type response observed in the degenerate doping regime, while shorter confinement lengths are found to lead to an overall broadening of this MIR band and a shift of the peak maximum to larger wavenumbers.}
		\label{fig2}
\end{figure}

In Figure \ref{fig2}a we have thus reproduced three spectra measured further into the FIR spectral range at low, moderate and degenerate doping levels after converting absorbance to optical conductivity $\sigma(\tilde{\nu}) = \sigma_1+i\sigma_2$ using the thin-film approximation and the Kramers-Kronig relation \cite{Zhang2013}. The spectrum from the intrinsic nanotube sample, $\sigma_i$, is here used as reference to calculate the difference signal $\Delta\sigma=\sigma_i-\sigma_d$ from the spectrum of the doped sample $\sigma_d$.

Next, these spectra can be analyzed quantitatively using the Drude-Smith single-scattering approximation:
\begin{equation}
\sigma(\tilde{\nu})_{DS}= \frac{\sigma_0}{1-i\cdot2\pi\,c\tilde{\nu}\tau}\left(1+\frac{b}{1-i\cdot 2\pi\,c\tilde{\nu}\tau}\right),
\label{eq:DS}
\end{equation}

\noindent
where $\sigma_0$ is the DC conductivity, $\tau$ is the average scattering time, $c$ is the speed of light and the parameter $b$ is interpreted as taking into account memory or 'persistence of velocity' during scattering events~\cite{Smith2001}. The observed increase of the real part $\Re \sigma$ with wavenumber (black data-points in Fig~\ref{fig2}a) and slightly negative values of $\Im \sigma$ (blue data-points) at low and moderate doping, indicate that optimal agreement with experiment can only be obtained for negative values of the $b$ parameter, here $\approx -0.65$ to $-0.75$. This indicates a predominance of backscattering, which, in nanoscale systems is sometimes attributed to carrier confinement \cite{Cocker2017}.

Discrepancies between the fit and experimental data around $1200\,\rm cm^{-1}$ are mostly due to fine structure in these spectra arising from vibrational Fano resonances that will be discussed later.

In the degenerate doping regime, the scattering parameter $b$ approaches zero (see fit in the right panel of Fig. \ref{fig2}a). This suggests that scattering appears to be essentially devoid of any memory effects. The resulting behavior of eq. \ref{eq:DS} thus corresponds to the Drude-Sommerfeld limit of the Drude-Smith model describing free carrier dynamics in the presence of momentum randomizing phonon-scattering \cite{Sommerfeld1928}. The fit corresponds to a DC conductivity of $400\,\rm S\,cm^{-1}$ and a scattering time $\tau$ of $20\,\rm fs$.

This fit can be used for estimating the intratube DC mobility, $\mu_0 = e\tau/m^*$, which becomes $500\,\rm cm^2\,V^{-1}\,s^{-1}$. Here we used a hole mass $m^*$ of 0.07 times the free electron mass \cite{Hartleb2015}. By comparison, mobilities observed in thin-film field-effect-transistor (FET) devices are more likely limited by intertube transport and thus typically feature values up to about $20\,\rm cm^2\,V^{-1}\,s^{-1}$~\cite{Brohmann2018, Schiessl2017, Bottacchi2015, Rother2017}. The mobility reported here thus compares better with mobilities for short channel length FETs, where conductance is less affected by intertube transport and becomes limited by diffusive intratube transport to a few hundreds of $\rm cm^2\,V^{-1}\,s^{-1}$ \cite{Brady2014}.

Similarly good agreement between experimentally measured intraband absorption and predictions of the Drude-Sommerfeld model was found for metallic and degenerately doped SWNTs~\cite{Zhukova2017,Gorshunov2018} as well as for graphene, the 2D parent material \cite{Mak2008, Mak2012, Lapointe2017, Horng2011}. However, fit parameters obtained from the Drude-Smith analysis of experimental data above are lacking a clear physical interpretation, specifically regarding the mechanistic origin of the strong backscattering indicated by negative values of the $b$ parameter. We have thus carried out Monte-Carlo simulations of 1D charge carrier transport to test whether the reported parameters are consistent with the previously discussed confinement of charge carriers in the low and moderate doping regimes \cite{Eckstein2017, Eckstein2019}. 

\begin{figure}[htbp]
	\centering
		\includegraphics[width=8.5 cm]{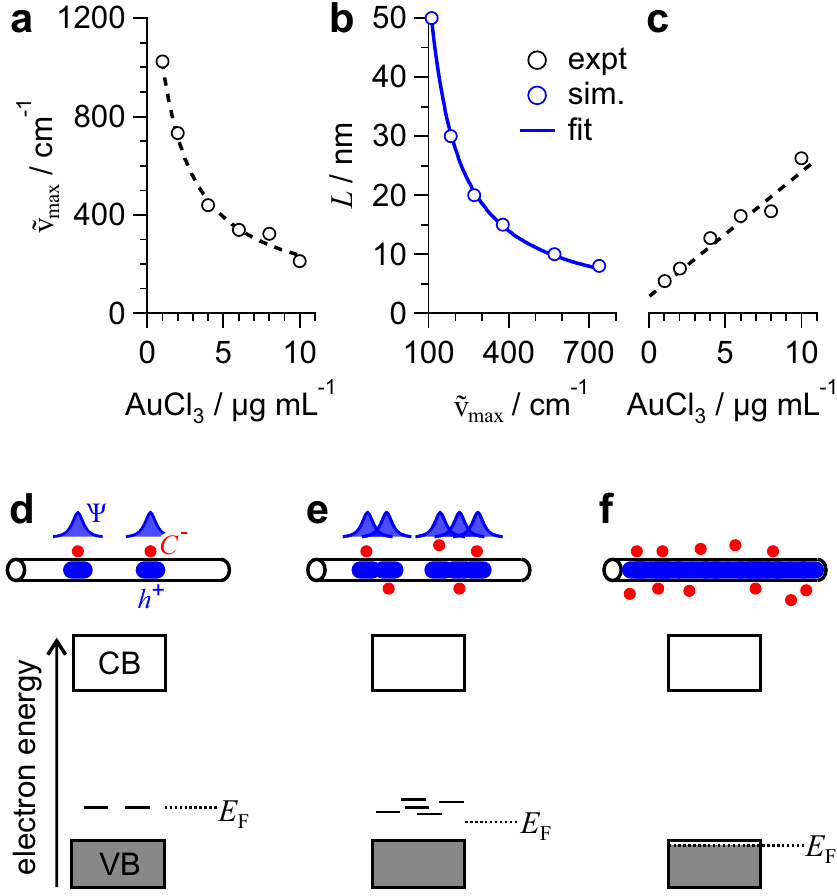}
		\caption{{\bf a)} Experimentally determined MIR peak positions as a function of doping level, here represented by the gold(III) chloride concentration used for doping. {\bf b)} Correlation of the confinement length $L$ used in Monte-Carlo simulations and the MIR peak maximum in simulated spectra. {\bf c)} Combination of data from {\bf a)} and {\bf b)} to obtain a correlation of dopant concentrations with the predicted $L$. {\bf d)} Schematic illustration of charge carrier localization by negative counterions $C^-$ for weakly doped s-SWNTs, {\bf e)} coalescence of impurity levels and emergent carrier delocalization within impurity bands at higher doping levels, and {\bf f)} complete delocalization of carriers in the degenerate doping regime.}
		\label{fig3}
\end{figure}

We incorporated the effect of confinement in nanotubes, following the approach by Cocker \textit{et al.} for transport in a 2D model Drude gas ~\cite{Cocker2017} by confining carrier trajectories to 1D segments of length $L$. Carrier motion within these segments is assumed to be free and backscattering occurs only if a trajectory reaches a segment boundary which is assumed to be perfectly reflective (see supporting information for simulation details).

Figure \ref{fig2}b shows simulations of the optical conductivity for different confinement lengths $L$ using the experimentally obtained scattering time $\tau=20$\,fs and the reported effective carrier mass $m^*=0.07\,m_e$ \cite{Hartleb2015}. As seen from the left panel of Figure \ref{fig2}b, simulations perfectly match predictions of the Drude model expected in the limit of very long segment lengths.

Deviations from the Drude model first become evident at small wavenumbers for confinement lengths of $L=200$\,nm or less (see center panel of Figure \ref{fig2}b), corresponding to typical nanotube lengths in our samples. These changes become more pronounced at even smaller confinement lengths, as shown in terms of the real part of the optical conductivity in the right panel of Figure \ref{fig2}b.

The results of these simulations are summarized in Figure \ref{fig3} where we compare the experimental absorbance peak positions at different doping levels (Figure \ref{fig3}a) with peak positions obtained from the simulations for different confinement lengths (Figure \ref{fig3}b). The product of the simulated peak position, $\tilde \nu_{\rm max}$, with the corresponding confinement length, $L$, is found to be constant with $L \tilde{\nu}_\text{max}=5570\,\rm cm^{-1}\,nm$. As shown in Figure \ref{fig3}c, the results from Figures \ref{fig3}a and \ref{fig3}b can be combined to obtain the dependence of the predicted confinement length on the dopant concentration. This behavior implies that charge carriers are localized to regions whose size increases with doping level. 

A linear extrapolation of the localization length in Figure \ref{fig3}c to very small doping levels yields a confinement length of 3.4\, nm. This is in excellent agreement with a previous estimate of $4\, \rm nm$. This value corresponds to the approximate spatial extent of calculated impurity level wavefunctions\cite{Eckstein2017} when stabilized by adsorbed chlorine counterions from the redox reaction used for doping \cite{Kim2011, Eckstein2017, Eckstein2019}. It is also consistent with the doping-induced bleach of the $X_1$ exciton band in Figure \ref{fig1}a, when sections of the nanotube occupied by such impurity states are assumed to no longer support an excitonic band-structure. 

The apparent increase of the confinement length with doping level in Figure \ref{fig3}c can now be explained using the schematic illustrations of Figure \ref{fig3}d-f. In the dilute case shown in Figure \ref{fig3}d, surplus charges are confined by impurity states $\Psi$. As mentioned above, these are bound by interaction with external counterions $C^-$. As shown schematically in Figure \ref{fig3}e, increased doping will then lead to the overlap of impurity wavefunctions, allowing charges to extend their excursion range progressively. In the degenerately doped case illustrated schematically in Figure \ref{fig3}f, excess carriers can move easily throughout the entire system, leading to the Drude gas-like optical response discussed above. 

Lastly, we note that a comparison with localized polaron states in the spectra of doped semiconducting polymers shows some similarities to the MIR features observed here. Polaron states are also believed to be localized by interaction with external counterions. The $\rm P_1$ polaron band maxima are then identified with the corresponding impurity ionization level \cite{Bredas1985}. If we use the same approach in combination with the measured MIR peak maximum in dilutely doped s-SWNTs at $\approx 1{,}200\rm \, cm^{-1}$, we would obtain a binding energy on the order of $150\,\rm meV$. This again, appears to be in good agreement with the impurity level binding energy for doped (6,5) s-SWNTs predicted by simple quantum chemical calculations \cite{Eckstein2017}.

\subsection{Fano resonances}

\begin{figure}[htbp]
	\centering
		\includegraphics[width=8.5 cm]{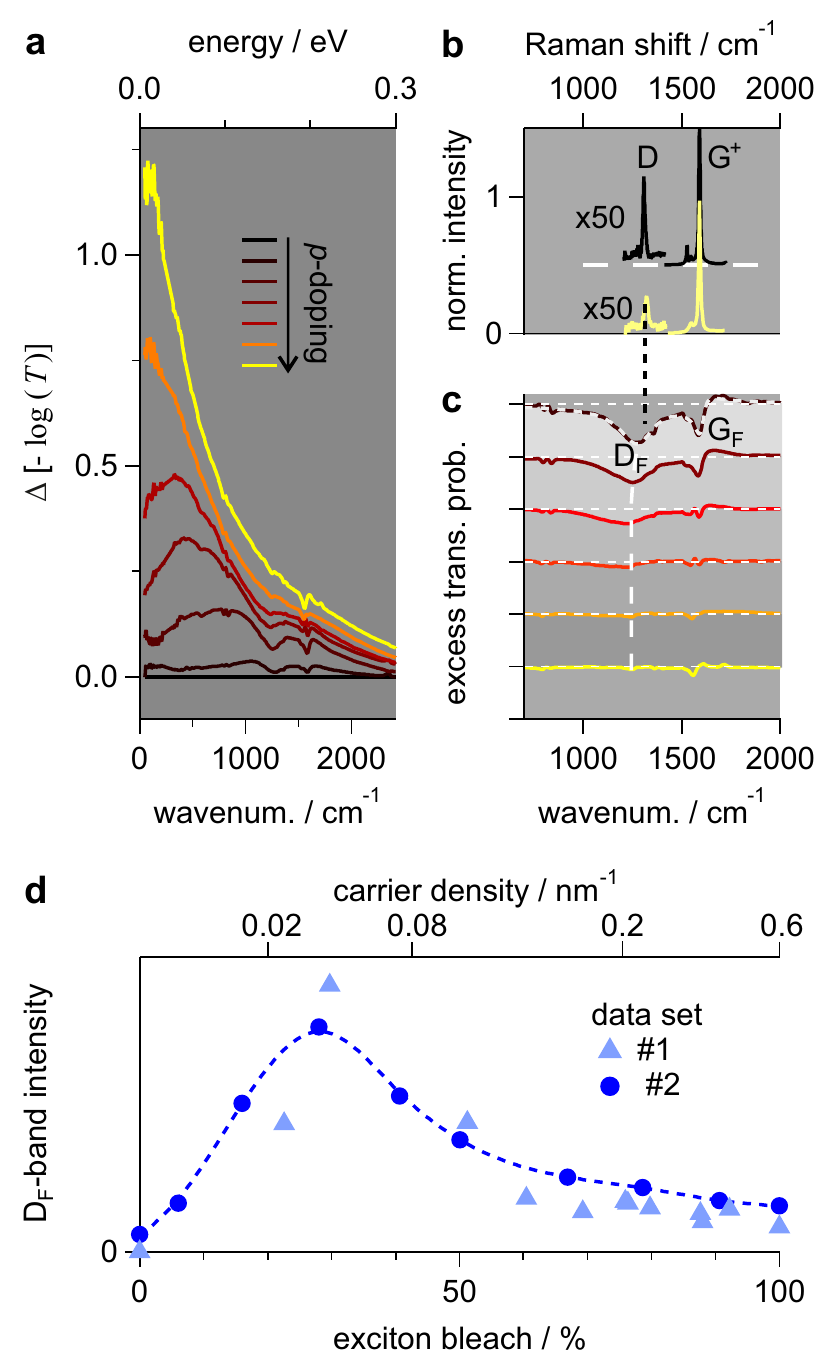}
		\caption{{\bf a)} Close up view of the FIR/MIR spectral region for increasing doping levels. {\bf b)} Raman D and $\rm G^+$ bands with the D band range expanded for clarity as reference for {\bf c)} the features seen in the excess transition probability for the same samples with increasing doping levels, weakly doped on top and degenerately doped at the bottom. The dashed line overlaid on the topmost spectrum is a two component Fano fit as described in the main text. {\bf d)} Dependence of the D band intensity on the doping level (top axis) as well as exciton bleach (bottom axis). The increase of the D band intensity up to about $0.03\rm\, nm^{-1}$ is attributed to an increasing concentration of localized carriers and scattering centers, while the eventual decrease at higher doping levels is due to charge carrier delocalization.}
		\label{fig4}
\end{figure}

In addition to the broad FIR/MIR response of the confined Drude gas shown in Figure \ref{fig4}a, FIR/MIR spectra also feature two antiresonance like absorption minima near to the nominal frequencies of the Raman $\rm G^+$-mode and the disorder induced D-mode of $1{,}589$ and $1{,}308\,\rm cm^{-1}$, respectively. 

To illustrate the evolution of these antiresonances with changing doping level more clearly, we calculate the so-called excess transition probability \cite{Fano1961} as quotient of the measured FIR/MIR signal and the unperturbed Drude-type background absorption. The latter is here estimated using a Gaussian fit to the high frequency falling edge of the Drude peak, while excluding the frequency range with the two prominent vibrational antiresonances by a zero-weight data mask. The resulting quotient $I_{\rm expt}/I_{\rm ref}$ is shown for different doping levels in Figure \ref{fig4}c. The series of spectra from low to high doping levels (black to yellow) reveals that the antiresonance excess transition probability is most strongly reduced in the weakly doped system (top brown curve). As a point of reference we also include Raman spectra in Figure \ref{fig4}b, corresponding to the weakly and degenerately doped system, with the intensity in the D-band range multiplied for clarity by a factor of 50. 

Bermudez and Ericson \cite{Bermudez2006} as well as Lapointe \textit{et al.}\cite{Lapointe2012}~have reported similar features in covalently functionalized nanotube samples of mixed chirality as well as in functionalized graphene \cite{Lapointe2017}. Lapointe \textit{et al.}~attributed these bands of increased transmissivity to Fano resonances brought about by the coupling of discrete vibrational modes to the continuum of electron-hole excitations in doped nanotubes. 

A fit of the topmost MIR dataset of Figure \ref{fig4}b using two Fano profiles of the type $(q+2\Delta \tilde\nu/\Gamma)^2/(1+(2\Delta \tilde\nu/\Gamma)^2)$ indeed provides a good description of the measured spectrum (see dashed line). $q$ here designates the Fano shape parameter, $\Delta \tilde\nu$ represents the wavenumber-shift with respect to the discrete mode and $\Gamma$ describes the strength of the coupling between the discrete mode, and the continuum. Peak positions obtained from this fit for weakly doped nanotubes yield the $\rm G_{F}$- and $\rm D_{F}$-Fano bands centered at $1{,}603$ and $1{,}280\,\rm cm^{-1}$ respectively. The corresponding coupling constants are $\Gamma_{\rm G_F}\approx 50\,\rm cm^{-1}$ and $\Gamma_{\rm D_F}\approx 200\,\rm cm^{-1}$. These widths are considerably larger than the widths of $9\,\rm cm^{-1}$ and $15\,\rm cm^{-1}$ observed for the Raman $\rm G^+$ and $\rm D$ modes, respectively. Moreover, the relative antiresonance oscillator strengths also appear to show distinctly different behavior if compared with Raman modes, where the $\rm G^+$-band is dominant.

A comparison of Raman and MIR Fano antiresonances thus suggests that the latter may arise from distinct albeit related scattering mechanisms. It is noteworthy, that such antiresonances were also reported for covalently functionalized and doped graphene \cite{Lapointe2017} as well as for functionalized larger diameter nanotubes \cite{Lapointe2012}, with surprisingly similar shape and intensity ratios. We thus propose that the mechanisms leading to the formation of MIR antiresonances in covalently functionalized graphene and redox-chemically doped SWNTs are closely related. 

\begin{figure*}[htbp]
	\centering
		\includegraphics[width=17.0 cm]{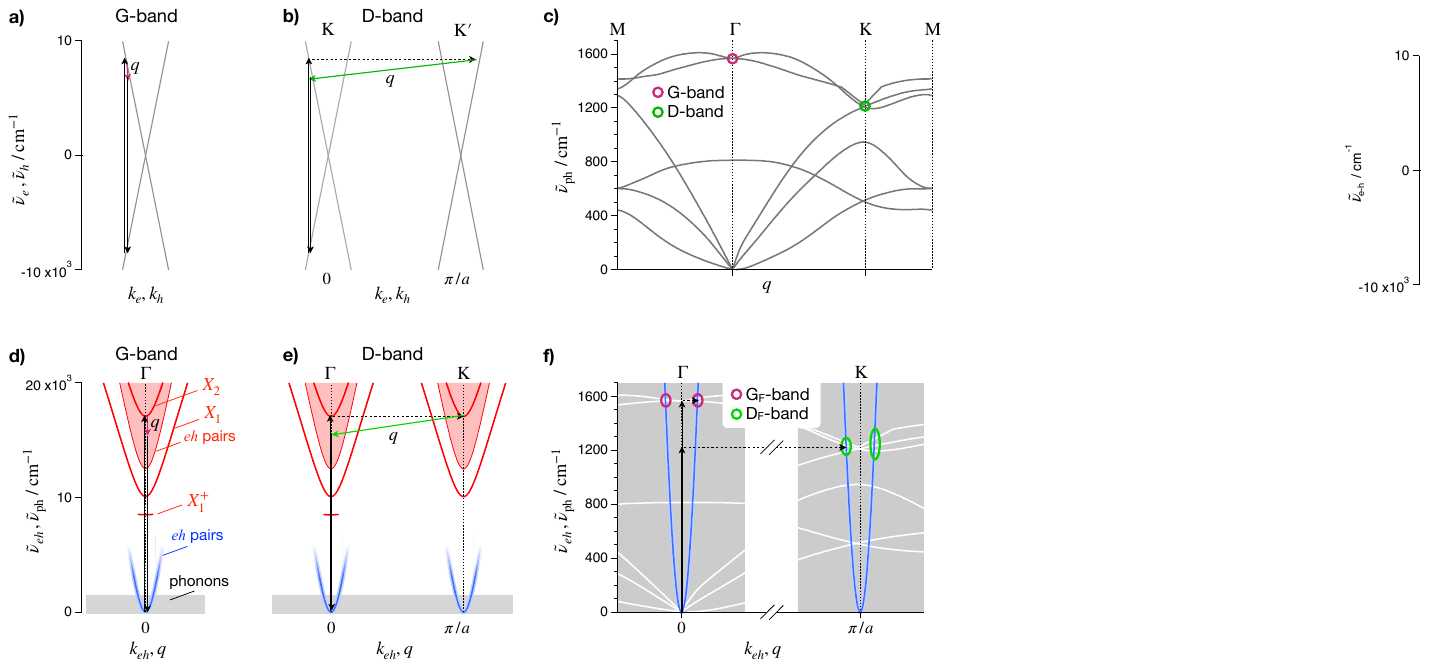}
		\caption{{\bf a)} Schematic illustration of the G-band Raman resonance in Graphene. {\bf b)} The same for the D-band intervalley double resonance with only one of the ‘incident’ resonances shown for simplicity. {\bf c)} Graphene phonon band structure as a reference. {\bf d)} Schematic illustration of $\rm G^+$ Raman resonance scattering for a p-doped semiconducting SWNT (for simplicity a (8,0) SWNT). Interband excitations such as excitons, trions and free e-h pairs are shown in red, intraband e-h pairs are shown in blue, and the region of the phonon spectrum is shown as grey area. {\bf e)} Extension of this scheme to the D-band double resonance. This schematic also includes intervalley excitations at the K point of this band structure. For simplicity the energies of these intervalley excitations at the K point are taken to be the same. The intraband e-h pair excitations at K (also shown in blue) are, for example, associated with e-h pairs with a hole in the single particle valence band at K (not shown) and an electron in the single particle valence band at K’ (also not shown). {\bf f)} Schematic of the mechanism leading to MIR Fano resonances with a magnified view of two regions in the phonon spectrum around the $\Gamma$ and K points. The purple and green ovals indicate where intraband e-h pair excitations (blue) come into resonance with phonon modes. Coupling of these coupled states to the incident MIR photons is then facilitated in a similar manner as for the Raman double resonance by defect scattering, indicated by the dashed horizontal lines.}
		\label{fig5}
\end{figure*}

To clarify this, we briefly review the resonant scattering mechanism leading to the formation of graphene $\rm G$ and $\rm D$ bands as shown in Figure \ref{fig5}a and \ref{fig5}b. The $\rm G$ band arises from an intravalley scattering process such as the one schematically illustrated for the incident photon resonance in Figure \ref{fig5}a. The $\rm D$ band on the other hand arises from the intervalley double resonance process where a lattice defect provides the momentum required to elastically access the $\rm K’$ valley (dashed line), here again shown only for one of the incident photon resonances \cite{Maultzsch2001}. The schematic phonon band structure in Figure \ref{fig5}c shows the emitted phonon states at the $\Gamma$- and $\rm K$-points of the graphene Brillouin zone.

The same types of scattering processes are also responsible for Raman resonances in semiconducting SWNTs. We here illustrate this in Figures \ref{fig5}d and \ref{fig5}e using a schematic of the dispersion of several types of excitations in a weakly doped (8,0) SWNT. Using the simpler band structure of an (8,0) -- instead of a (6,5) s-SWNT -- here merely serves to reduce the complexity of their representation. The fundamental mechanisms are not affected by this simplification. Figure \ref{fig5}d includes interband free and bound electron-hole pair excitations as well as trions (in red) and intraband electron-hole pair excitations (in blue). The range of the graphene phonon band structure is indicated by the grey shaded region. Intervalley excitations in Figure \ref{fig5}e are expected to have slightly different energies if compared with intravalley excitations, but for simplicity are shown at the same level. 

Figure \ref{fig5}f represents enlarged sections near $\Gamma$ and $\rm K$ points of the phonon energy range of Figure \ref{fig5}e. Here, the spectrum of intraband electron hole pair excitations is seen to come into resonance with some of the optical phonon branches in the vicinity of the $\Gamma$ and $\rm K$ points. These points are marked by purple and green ovals. Zone folding of the graphene phonon bands may lead to some additional dispersion of the accessible phonon modes but without modifying the character of the resonant process. The photon momentum however is too small to allow access to these resonances such that scattering by a defect must make up for the remaining momentum mismatch, as indicated in Figure \ref{fig5}f, again using dashed horizontal lines.

The observation of such resonances in the redox-doped s-SWNTs thus provides further evidence for the confinement and localization of excess charge carriers by impurity potentials. Without such confinement it would be difficult to reconcile the proposed scattering process with momentum conservation.

The observation of Fano resonances in this system thus appears to indicate that the SWNT symmetry is broken by impurity potentials, recently attributed to the interaction with external counterions \cite{Eckstein2017, Eckstein2019}. The strength of these features should thus provide some insight into the strength of the impurities scattering potential. In Figure \ref{fig4}d we thus plot the intensity of the $\rm D_F$ band as a function of the exciton bleach and the corresponding previously estimated carrier density \cite{Eckstein2019}.

Both datasets shown in Figure \ref{fig4}d indicate that the intensity of the $\rm D_F$-band increases linearly up to a carrier density of about $0.03\,{\rm nm^{-1}}$ or one charge impurity per about $30\,{\rm nm}$ of tube length. This suggests that added hole states continue to be localized throughout the weakly-doped regime. Note, that the Raman D to G-band intensity ratio decreases over this doping range and shows no similar intensity maximum as the MIR $\rm D_F$-band. 

The following decrease of the $\rm D_F$-band intensity at higher doping levels is here attributed to an increasing probability for coalescence of impurity levels as evidenced by the overall red-shift of the broad FIR/MIR peak maximum discussed in the previous paragraphs. In the degenerately doped system the $\rm D_F$-band intensity once again vanishes, in accordance with expectations for a homogeneous distribution of charge carriers, which restores the initially broken symmetry. These findings are also consistent with the absence of pronounced phonon antiresonances in the spectra of metallic nanotubes despite the existence of a similar low-energy electronic continuum absorption~\cite{Lapointe2012, Kampfrath2008, Pekker2006, Borondics2006}.

\section{Summary and Conclusions}

We have studied the infrared response of doped (6,5) carbon nanotubes over a wide range of doping levels using broadband transmission spectroscopy. The doping-induced bleach of interband exciton transitions in the NIR is accompanied by a proportionate increase of a Drude-type intraband absorption signal in the MIR. However with increasing doping level, its spectral weight shifts to {\it smaller} energies, defying general assumptions with regards to the plasmonic nature of this band. This behavior is taken as evidence for carrier confinement in the low and moderate doping regimes. Surprisingly simple Monte-Carlo simulations of frequency-dependent carrier transport, allow to estimate the confinement length of carriers at very low doping levels of about $3.4\, \rm nm$. This is in good agreement with earlier reports \cite{Eckstein2017} and suggests that the fundamental physics of the Drude band formation is captured by this simple model. For the degenerately doped system the MIR response no longer shows any evidence of carrier confinement.

Moreover, carrier confinement was found to be a prerequisite for the observation of strong MIR, impurity-induced $\rm D_F$ and $\rm G_F$ Fano antiresonances near to the corresponding Raman bands. A preliminary estimate of the coupling strength between e-h pair excitations and vibrational modes based on a two component Fano fit yields $\Gamma_{\rm G_F}\approx 50\,\rm cm^{-1}$ and $\Gamma_{\rm D_F}\approx 200\,\rm cm^{-1}$. The mechanism underlying the formation of both Fano bands appears to be of second order as it involves scattering with an impurity in a manner similar to the formation of the Raman D-mode double resonance. The process is based on the resonance between electron hole pair excitations and one or several phonon modes as well as with the incident electromagnetic radiation by virtue of impurity scattering. The disappearance of these antiresonances at higher doping levels is consistent with carrier delocalization in the degenerately doped regime. 

Similarities of shape and relative intensities of both $\rm G_F$ and $\rm D_F$ resonances  with those found for covalently functionalized large diameter SWNTs as well as for functionalized graphene are intriguing. Specifically since we would expect the different character of impurities in covalently functionalized lattices and in the redox doped nanotubes studied here, to lead greater differences between the associated MIR Fano Resonances \cite{Bermudez2006, Lapointe2012, Lapointe2017}.

The findings presented here confirm that the mid-infrared response of doped s-SWNTs is rich with information about carrier localization and delocalization as well as the coupling of electronic and lattice degrees of freedom.

\begin{suppinfo}
Control experiments, further experimental spectra, description of the Monte Carlo simulations.
\end{suppinfo}

\section{Acknowledgements}
All authors acknowledge financial support by the German National Science Foundation through the DFG GRK2112. T.H. and K.E. also acknowledge financial support through DFG grant HE 3355/4-1. K.E. acknowledges assistance by Florian Oberndorfer and Michael Auth for the fabrication of nanotube suspensions and the determination of nanotube film thicknesses. All authors acknowledge support by I. Fischer for providing access to crucial experimental infrastructure. RM acknowledges Financial support from the Natural Sciences and Engineering Research Council of Canada (NSERC) under Grant Nos. RGPIN-2019-06545 and RGPAS-2019-00050 and the Canada Research Chair programs.

\end{document}


\newpage
\section{Control Experiments}

\begin{figure}%
	\includegraphics[width=16.5cm]{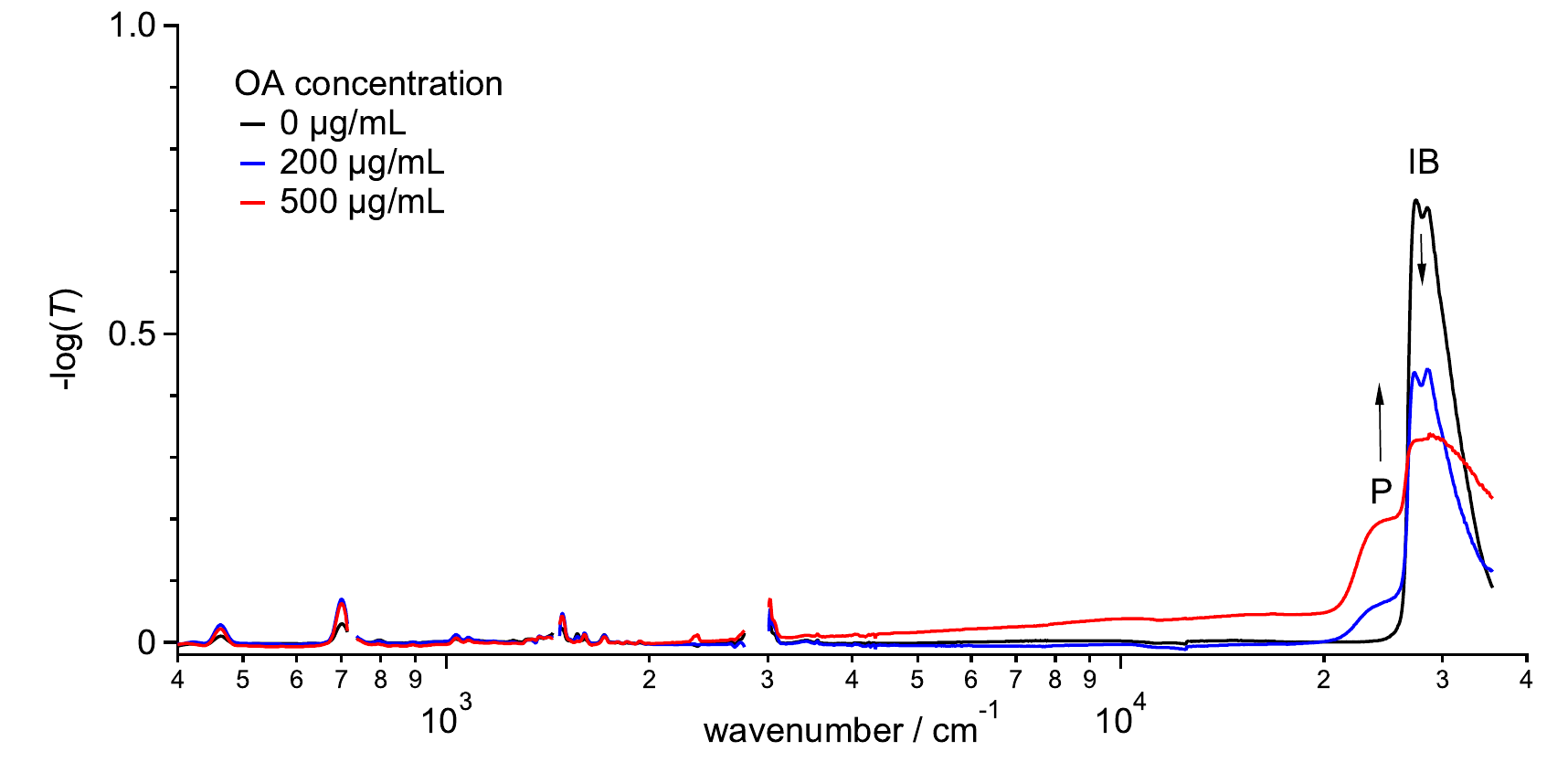}%
	\caption{Transmission spectra of an intrinsic and doped PFO-BPy polymer film.}%
	\label{SI_Fig1}%
\end{figure}

Figure~\ref{SI_Fig1} shows transmission spectra of a PFO-BPy film (obtained by dropcasting a PFO-Bpy solution in toluene on a 500\,µm thick PE substrate) after treatment with OA solutions (volume ratio toluene:acetonitrile 5:1) of different concentration.
The undoped sample (black spectrum) shows strong HOMO-LUMO or interband (IB) absorption at $\approx 3.45$\,eV and minor absorption in the infrared due to vibrational transitions.
Doping with OA causes a decrease of the IB band and an increase of the polaron band P at $\approx 3.0$\,eV but no broad electronic continuum absorption in the infrared.
Slight intensity variations of the vibrational bands probably arise because of measuring different positions of the inhomogeneously thick polymer film obtained by dropcasting.

\begin{figure}%
	\includegraphics[width=16.5cm]{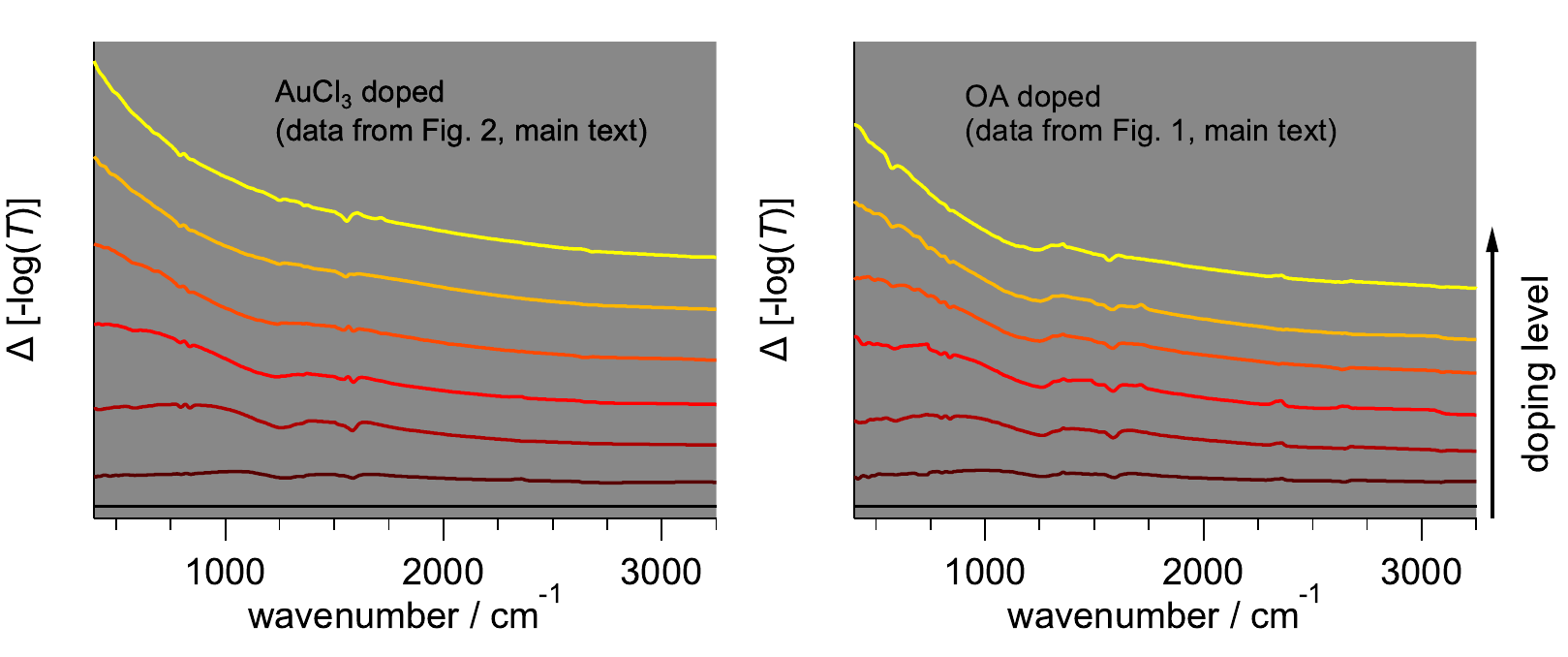}%
	\caption{Difference spectra using AuCl$_3$ or OA as hole dopant.}%
	\label{SI_Fig2}%
\end{figure}

Significant contributions from the OA and AuCl$_3$ dopants and their reaction products to the IR spectra of doped SWNT films can be excluded, since both dopants give rise to similar spectra, as shown in Figure~\ref{SI_Fig2}.

\section{Further Experimental Details}

\subsection{Data recording and processing}
Film samples were fixed in the light path of the spectrometers using tape and an appropriate sample holder. 
All spectra were background corrected using reference spectra of the pure substrate.
To cover the large spectral range from the far infrared to the UV, three different spectrometers were used (see main text).
Transmission spectra in the mid-infrared ($\geq 400$\,cm$^{-1}$) were mainly recorded at a Jasco FT/IR-
4100 spectrometer using a high-intensity ceramic IR light source, a Ge/KBr beam splitter and a triglycine sulfate detector.
The KBr beam splitter transmittance drops severely in the far infrared at wavenumbers $< 400$\,cm$^{-1}$. Therefore, transmission spectra in the far infrared were recorded using an Bruker IFS120HR spectrometer with a globar light source, a Mylar beam splitter and a triglycine sulfate detector.
The three individual spectra of each sample were concatenated using regions of spectral overlap for mutual adjustment by using small constant offsets.

\subsection{Absorption spectra of the substrates}
\begin{figure}
	\includegraphics[width=16.5cm]{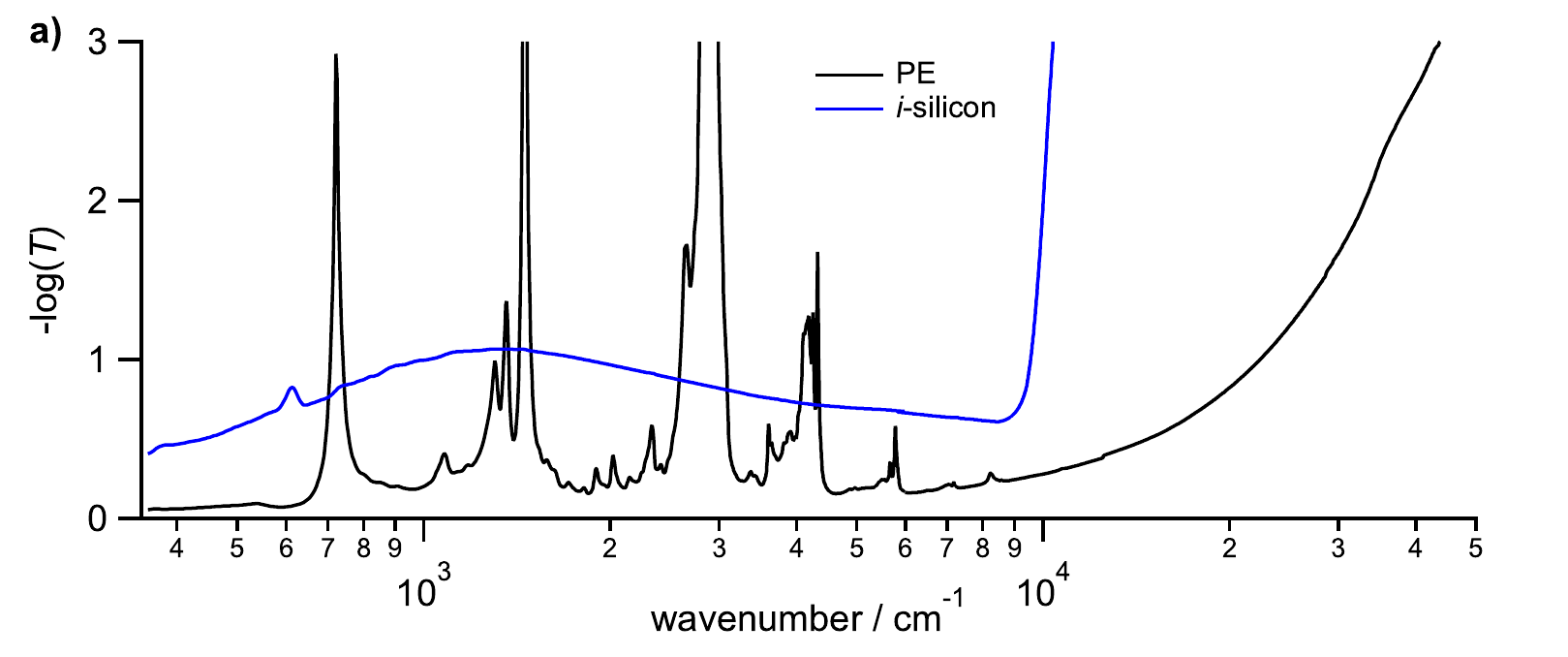}%
	
	\includegraphics[width=16.5cm]{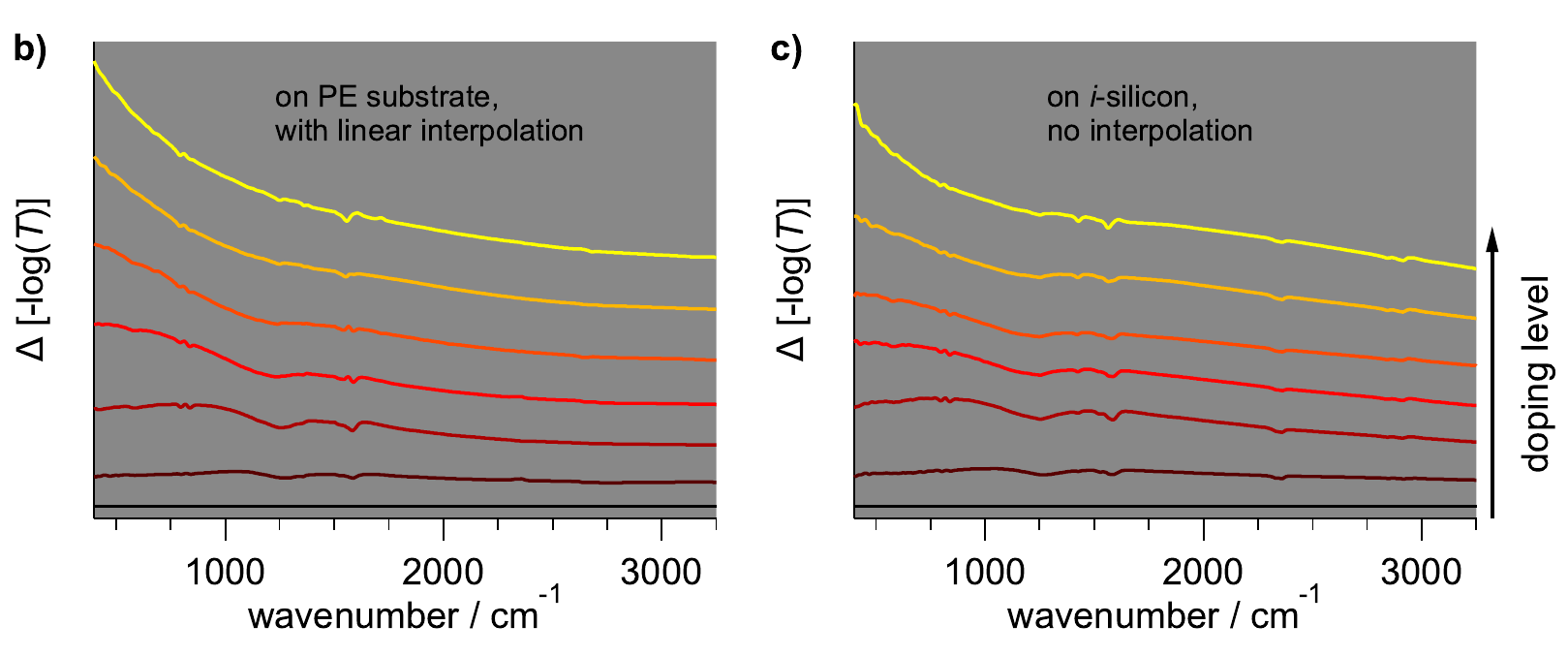}%
	\caption{a) Broadband transmission spectra of the PE and intrinsic silicon substrates. b), c)~Comparison of difference spectra using both substrates with doping level increasing from bottom to top.}%
	\label{SI_Fig3}%
\end{figure}

Absorption spectra of the PE and intrinsic silicon substrates are shown in Figure~\ref{SI_Fig3}a.
500\,µm thick PE substrates were chosen to cover the broad spectral range from the far infrared to the UV being almost fully transmissive and thus
allowing to measure electronic intra- and interband transitions on the same SWNT sample.
In the narrow spectral range where the PE substrate is opaque, the SWNT film spectra were linearly interpolated.
The introduced error was found to be small as checked by comparison to SWNT film samples on intrinsic silicon (see Figure~\ref{SI_Fig3}b and c)
\clearpage
\subsection{Difference spectra from the main text over a larger spectral range}
\begin{figure}%
	\includegraphics[width=16.5cm]{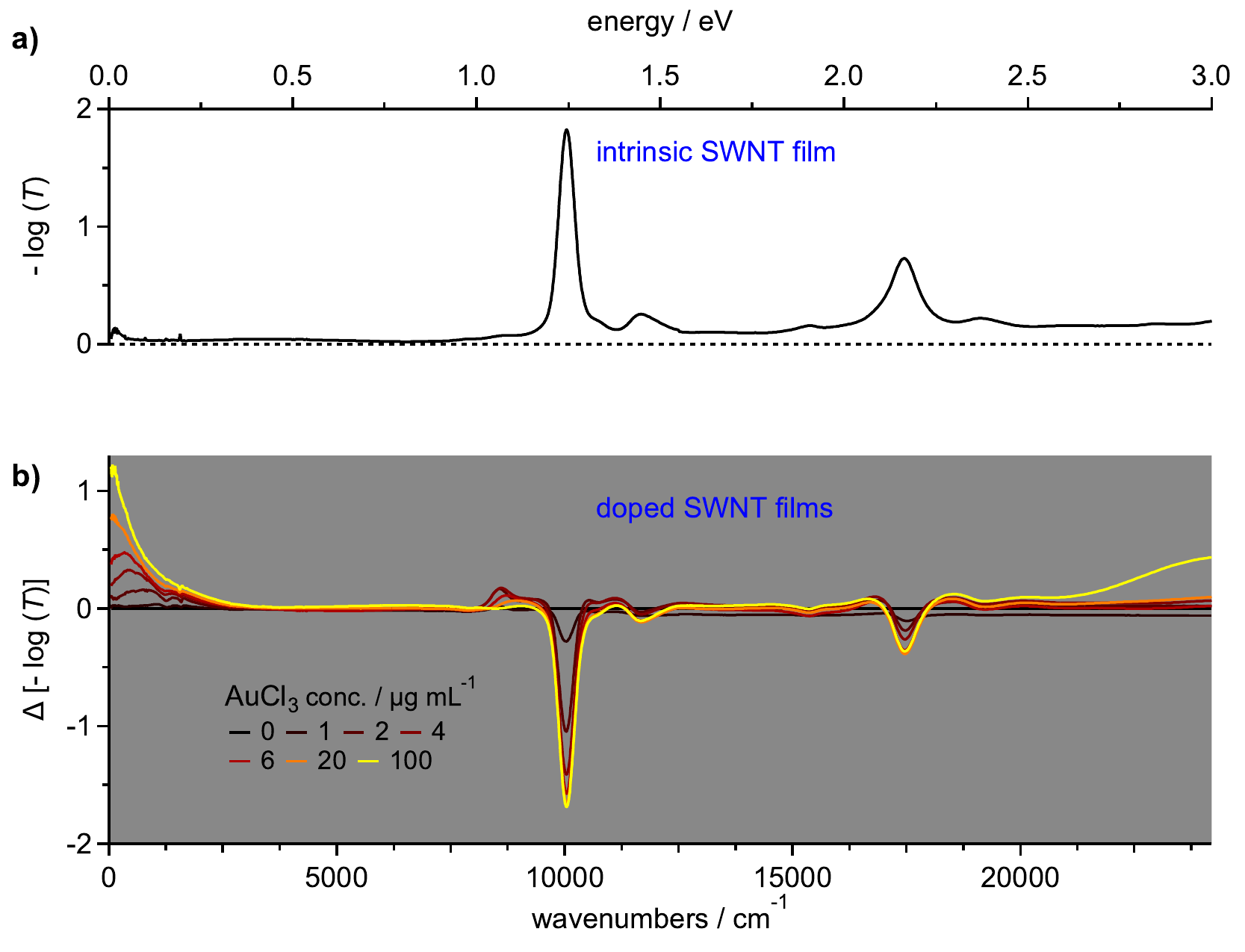}%
	\caption{a) Broadband spectrum of an intrinsic SWNT film and b) difference spectra after AuCl$_3$ treatment.}%
	\label{SI_Fig4}%
\end{figure}

SI Figure~\ref{SI_Fig4}b shows the difference spectra of Figure~4a of the main text over a larger spectral range.
SI Figure~\ref{SI_Fig4}a illustrates the spectrum of the intrinsic film which was used as a reference for the difference spectra.

\section{Calculation of the optical conductivity}
The broadband transmission spectra were converted into optical conductivity spectra by the Kramers-Kronig relation and
under the thin-film approximation as described by Zhang et al.~\cite{Zhang2013}.
The Kramers-Kronig transformation of the transmission data $T(E)$ resulted in the phase spectrum $\phi(E)$ via
\begin{equation}
\phi(E)=-\frac{E}{\pi}\int{\frac{\ln (T(E')) - \ln (T(E))}{E'^2-E^2} dE'}.
\label{eq:phaseSpec}
\end{equation}
Next, the phase spectrum was used to calculate the transmission coefficient $t(E)$
\begin{equation}
t(E)=\sqrt{T(E)}\exp(i\,\phi(E)),
\label{eq:transCoeff}
\end{equation}
from which the optical conductivity $\sigma(E)$ is obtained under the thin-film approximation by
\begin{equation}
\sigma(E)=\frac{(1+n)(1-t(E))}{t(E)\,Z_0\,d},
\label{eq:OptCond}
\end{equation},
where $n=1.5$ is the refractive index of the polyethylene substrate, $Z_0=377\,\Omega$ is the vacuum impedance and $d$ is the film thickness.

\begin{figure}%
	\includegraphics[width=16.5cm]{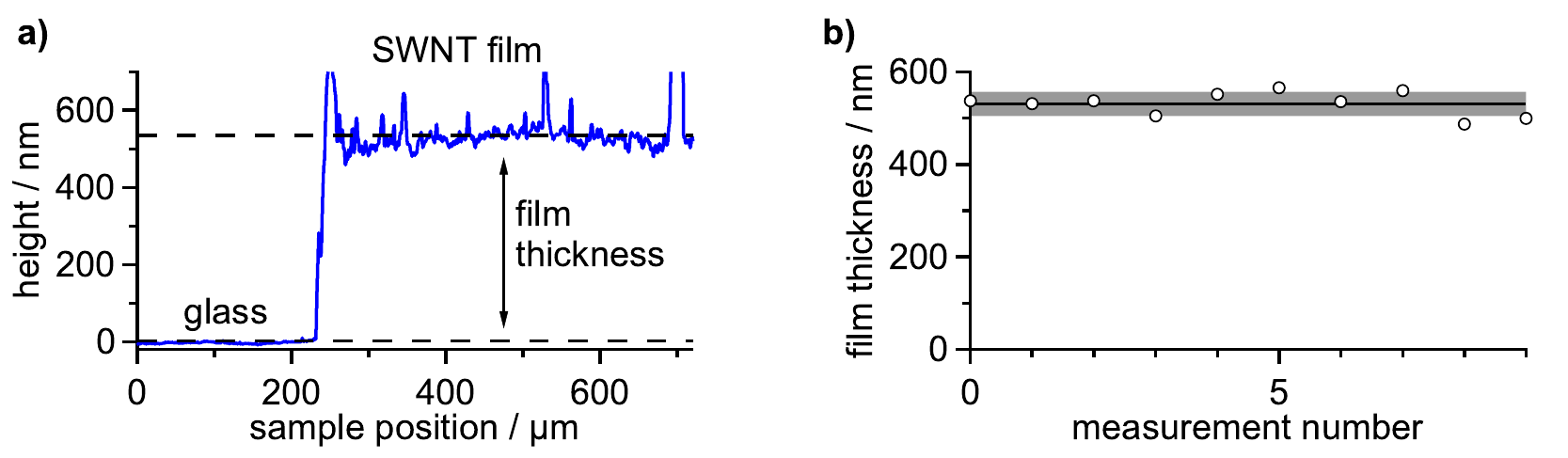}%
	\caption{a) Single measurement with the profilometer. b) Results of ten measurements yielding an average film thickness of $d=(531\pm26)$\,nm.}%
	\label{SI_Fig5}%
\end{figure}

The film thickness was determined using a Dektak 150 (Veeco Instruments Inc.) profilometer using identical films on a glas substrate with reduced surface roughness compared to polyethylene substrates.
Figure~\ref{SI_Fig5}a shows a single profilometer measurement whereas Figure~\ref{SI_Fig5}b shows the determined film thicknesses of ten measurements yielding an averaged thickness of $d=(531\pm26)$\,nm.

\section{Description of the Monte Carlo simulations}
\begin{figure}%
\includegraphics[width=16.5cm]{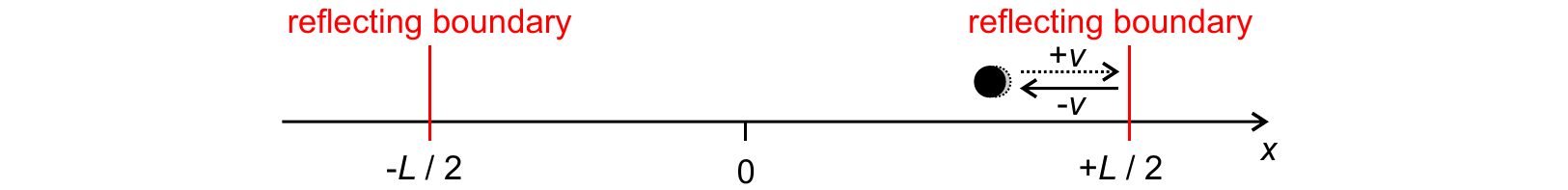}%
\caption{Motion of a charge carrier particle (black circle) confined to a 1D box of length $L$ and reflective boundaries (100\,\%  reflectivity). 
Once a boundary is hit, backscattering changes the particle velocity from $v$ to $- v$.}%
\label{SI_Fig6}%
\end{figure}

The simulations were performed as described by Cocker \textit{et al.} for two-dimensional charge transport~\cite{Cocker2017}.
In this publication, the same ideas were followed to describe charge transport in 1D.
Monte Carlo (MC) simulations are based on a large number of identical random experiments.
Here, the motion of a single charge carrier under the influence of an oscillating electric field is simulated.
This simulation is then repeated for a large number of particles $N_\text{part}$.
Charge carrier localization is introduced --~similar to Cocker \textit{et~al.}~-- by restricting carrier motion to a 1D box of localization length $L$ and perfectly reflecting boundaries (see Fig.~\ref{SI_Fig6}).
The following sections describe the MC algorithm used for the simulations.
\subsubsection{Initialization}
Each particle $p$ entering the simulation is described by two variables: the position $x$ and the velocity $v$ in $x$-direction.
At time $t=0$ the particle is initialized with a velocity $v_0$, randomly chosen from a Gaussian distribution with zero mean and $v^2_\text{th}$ variance.
Following the ideal gas theory, the thermal velocity $v_\text{th}=\sqrt{kT/m^*}$ is the root mean square value of the Gaussian velocity distribution .
The initial position $x_0$ is randomly taken with equal probability to be within $- L/2\leq x \leq +L/2$.
\subsubsection{Scattering processes}
Two different types of scattering events are taken into account by the simulations: velocity randomizing scattering and backscattering.

Within the relaxation time approximation, the probability $W$ for carrier scattering after a time interval $\Delta t$ is $W=\Delta t/\tau$.
In case of a scattering event, the velocity of the particle is again randomly reinitialized (as described above).
The time interval $\Delta t\ll\tau$ is here associated with the time step size of the simulation.

Backscattering at the reflecting boundaries at $x=\pm L/2$ is implemented such that
particles, which cross the boundaries during the time step $\Delta t$ change their velocity from $v$ to $-v$.
\subsubsection{Frequency dependence}
The dependence of the conductivity on the driving field frequency $f$ is obtained by separate simulations for each frequency. 
An oscillating electric field is applied and the particle velocity $v$ is evaluated at each time step $q$ according to
\begin{equation}
v(t=q\Delta t)=v_q=v_{q-1}+\frac{eE_0\Delta t}{m^*}\cos(2\pi f\times q\Delta t),
\label{eq:velocity}
\end{equation}
where $m^*$ is the effective carrier mass and $E_0$ is the maximum amplitude of the electric field.

The current density $J(t)$ in the time-domain is obtained by averaging the velocities of all $N_\text{part}$ individual particles $p$ via
\begin{equation}
	\langle v(t)\rangle=\frac{1}{N_\text{part}}\sum_p^{N_\text{part}}{v_p}(t). 
\label{eq:avVelocity}
\end{equation}
and is the simply given by $J(t)= n\langle v(t)\rangle e$ with the charge density $n=N_\text{part}/V$ and the unit volume $V$.

The frequency-dependent conductivity $\sigma(\omega)$ is given by: 
\begin{equation}
\sigma(\omega)=J(\omega)/E(\omega).
\label{eq:sigma_of_omega}
\end{equation}
This equation can be rewritten by expressing $J(\omega)$ and $E(\omega)$ by their Fourier transforms:
\begin{equation}
\sigma(\omega)=\frac{\int_{- \infty}^\infty{J(t)\exp(i\omega t)\dd{t}}}{\int_{- \infty}^\infty{E(t)\exp(i\omega t)\dd{t}}}.
\label{eq:sigma}
\end{equation}

A conversion of the integrals into sums is possible due to the discrete sampling:
\begin{equation}
\sigma(\omega)=\frac{\sum_{j=1}^T{J(t_j)[\cos(\omega t_j)+i \sin(\omega t_j)]}}{\sum_{k=1}^T{E_0 \cos(\omega t_k)[\cos(\omega t_k)+i \sin(\omega t_k)]}}.
\label{eq:sigma1}
\end{equation}
Here, $T$ is the total number of time steps in the simulation and the identity $\exp(i\omega t) = \cos(\omega t) + i\sin(\omega t)$ was used. 


Finally, $\sigma(\omega)$ is normalized to the DC value of the Drude model $\sigma_\text{0,D}=n e^2\tau/m^*$.
\subsubsection{Input parameters}
The average scattering time $\tau$ and the localization length $L_\text{loc}$ mainly determine the underlying physics described by the simulation.

Other input parameters, which mainly determe the accuracy and signal-to-noise ratio of the simulations, are the time step size $\Delta t$, the number of particles $N_\text{part}$ and time steps $T$ per driving frequency and the peak amplitude $E_0$ of the electric field.
Here, a time step size $\Delta t=0.2$\,fs is chosen.
The number of time steps is $T=10^{5}$ resulting in a total simulated time $T\Delta t=20$\,ps after the turn-on of the electric field. 
The driving frequencies are $f\geq (20\text{\,ps})^{- 1}= 0.05$\,THz, which ensures that all simulations contain at least one period of the driving frequency.
The number of particles used for each frequency is $N_\text{part}=8\times 10^4$, providing for a electric field maximum $E_0= 1$\,kV/ a reasonable compromise between signal-to-noise ratio and computation time.
The peak electric field $E_0= 1$\,kV/cm is clearly below the saturation field strength of $\approx 5-10\,$kV/cm reported for carbon nanotubes and approximately at the end of the linear response regime~\cite{Perebeinos2005,Perebeinos2006}.

\section{Correlation of the IR peak position and the localization length}

\begin{figure}%
	\includegraphics[width=8.5cm]{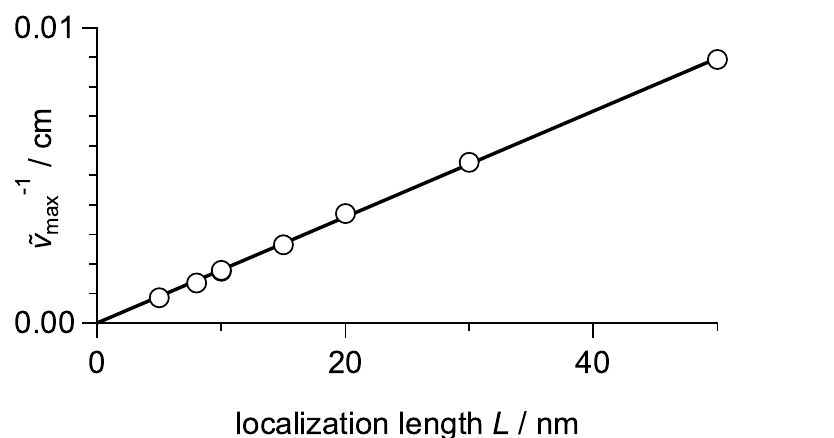}%
	\caption{Correlation of the inverse IR peak position and the localization length.}%
	\label{SI_Fig7}%
\end{figure}

The simulated real parts $\sigma_1$ for different localization lengths $L$ are shown in Figure~2b of the main text.
The spectral position of the peak maximum $\tilde{\nu}_\text{max}$ is obtained by fitting a Gaussian profile in the vicinity of the peak maximum.
Figure~\ref{SI_Fig7} shows the linear relationship between the inverse IR peak position $\tilde{\nu}_\text{max}^{-1}$ and the localization length $L$ used as input parameter for the Monte-Carlo simulations.
This linear relationship is expressed as $L=a\,\tilde{\nu}_\text{max}^{-1}$ with $a=5573\,\rm cm^{-1}\,nm$.

